\documentclass[12pt, preprint]{aastex}

\newcommand{\ha}{H$\alpha$}
\newcommand{\sm}{$\sim$}
\begin{document}

\title{Study of Rapid Formation of a $\delta$ Sunspot Associated with the 2012 July 2 C7.4 Flare Using High-resolution Observations of New Solar Telescope}

\author{Haimin Wang$^1$, Chang Liu$^1$, Shuo Wang$^1$, Na Deng$^1$, Yan Xu$^1$, Ju Jing$^1$, and Wenda Cao$^2$}

\affil{1. Space Weather Research Laboratory, New Jersey Institute of Technology, University Heights, Newark, NJ 07102-1982, USA;\\haimin.wang@njit.edu}

\affil{2. Big Bear Solar Observatory, New Jersey Institute of Technology, 40386 North Shore Lane, Big Bear City, CA 92314-9672, USA}

\begin{abstract}

Rapid, irreversible changes of magnetic topology and sunspot structure associated with flares have been systematically observed in recent years. The most striking features include the increase of horizontal field at the polarity inversion line (PIL) and the co-spatial penumbral darkening. A likely explanation of the above phenomenon is the back reaction to the coronal restructuring after eruptions: a coronal mass ejection carries the upward momentum while the downward momentum compresses the field lines near the PIL. Previous studies could only use low resolution (above 1\arcsec) magnetograms and white-light images. Therefore, the changes are mostly observed for X-class flares. Taking advantage of the 0$\farcs$1 spatial resolution and 15~s temporal cadence of the New Solar Telescope at Big Bear Solar Observatory, we report in detail the rapid formation of sunspot penumbra at the PIL associated with the C7.4 flare on 2012 July 2. It is unambiguously shown that the solar granulation pattern evolves to alternating dark and bright fibril structure, the typical pattern of penumbra. Interestingly, the appearance of such a penumbra creates a new $\delta$ sunspot. The penumbral formation is also accompanied by the enhancement of horizontal field observed using vector magnetograms from the Helioseismic and Magnetic Imager. We explain our observations as due to the eruption of a flux rope following magnetic cancellation at the PIL. Subsequently the re-closed arcade fields are pushed down towards the surface to form the new penumbra. NLFFF extrapolation clearly shows both the flux rope close to the surface and the overlying fields.

\end{abstract}


\section{INTRODUCTION}

Solar eruptive events, including flares, filament eruptions, and coronal mass ejections (CMEs) are due to magnetic reconnection or loss of equilibrium in the solar corona (see, e.g., a recent review by Webb \& Howard 2012). From time sequence magnetograph observations, irreversible and rapid changes of surface magnetic fields associated with a large number of major flares have been observed (e.g., Wang 1992, 2006; Kosovichev \& Zharkova 2001; Wang et al. 1994, 2002, 2004a,b; Liu et al. 2005; Sudol \& Harvey 2005; Petrie \& Sudol 2010; Burtseva \& Petrie 2013). Wang \& Liu (2010) synthesizes earlier studies and presented analysis of new events, and found a trend of increase of horizontal field at the polarity inversion line (PIL) associated with almost all the flares studied by the authors. The results agree with our previous finding of rapid changes of sunspot structure associated with flares (Liu et al. 2005; Deng et al. 2005), where the spot feature near the flaring PIL darkens while part of peripheral penumbra decays. Recently, the Helioseismic and Magnetic Imager (HMI) on board the Solar Dynamics Observatory (SDO) is providing state-of-the-art observations under seeing-free condition, which help advance the study of magnetic field changes associated with flares (e.g., Liu et al. 2012; Wang et al. 2012a,b; Petrie 2012). All these photospheric magnetic field changes are interpreted due to either the change of field line orientation or the appearance of newly formed magnetic loops near the surface. In both cases, the changes are more prominent in the horizontal rather than the vertical component.

From the viewpoint of the theory of the flare phenomenon, Hudson et al. (2008) and Fisher et al. (2012) pointed out that the coronal energy release and magnetic restructuring may cause back reaction to the solar surface and interior. The authors found that after a flare, the photospheric magnetic field may become more horizontal at the PIL. The authors used the simple principle of energy and momentum conservation, and specifically predicted  that flares can be accompanied by rapid and irreversible changes of photospheric magnetic field.  Melrose (1997, 2012) also provided an explanation for the enhancement of magnetic shear at the flaring PIL, using the concept of reconnection between two current-carrying flux systems. Such a magnetic share increase at the PIL is often observed (Wang et al., 2012a, b).   The work of Melrose and those of Hudson and Fisher, may have physics linkage: there are two current systems, one moves upward as part of an eruption and the other moves toward the surface. These might be linked to the tether-cutting reconnection model for sigmoids (Moore et al. 2001, 2006).

It is noticeable that even with the SDO/HMI observations, the resolution of data is no better than 1\arcsec. Therefore, the detailed evolution of sunspot structure may not be well observed. The high-resolution (0$\farcs$1) observation with the 1.6~m New Solar Telescope (NST; Goode et al. 2010; Cao et al., 2010) at Big Bear Solar Observatory (BBSO) thus provides a unique opportunity to understand the fine-scale structure change of sunspots associated with flares. Compared with most of the previous observations mainly for X-class flares, NST images also allows such a study related to weaker flares. In this Letter, we provide a detailed analysis of rapid formation of a $\delta$ sunspot associated with the C7.4 flare on 2012 July 2 from the NOAA AR 11525. We describe the multiwavelength observations and results in Section~\ref{result}, and summarize and discuss our major findings in Section~\ref{summary}. More dynamic detail can be seen in the accompanying animations in the online journal.

\section{OBSERVATIONS AND RESULTS} \label{result}

NOAA AR 11525 produced numerous C-class and M-class flares during its disk passage from 2012 June 29 to July 12. Figure~\ref{f1}(f) displays a HMI vertical field covering the entire active region. Multiple PILs and the possible interaction between emerging flux systems could be the primary reasons of the high flare productivity. We concentrate our analysis on a small field-of-view (FOV) (27\arcsec\ $\times$ 27\arcsec) as marked by the black box in (f), in order to scrutinize the flare-related changes of sunspot structure. This FOV includes a main flaring PIL, and is within the larger 70\arcsec\ $\times$ 70\arcsec\ FOV of NST diffraction-limited observation. NST best covers the C7.4 flare on July 2, which started at 18:45~UT, peaked at 18:56 UT, and ended at 19:02~UT in GOES 1--8~\AA\ flux. The related NST observations are at three wavelengths, the H$\alpha$ line center, H$\alpha -$0.75~\AA\ offband, and the TiO band at 7057~\AA\ which is a good proxy for the photospheric continuum. The cadence of H$\alpha$ blue-wing images is 6~s, while that of H$\alpha$ line center and TiO images is 15~s. All the observations were taken with the BBSO's 76 element adaptive optics, which is able to correct most of the atmospheric seeing at these wavelengths. Speckle reconstruction using 100 frames (obtained within 3~s) is subsequently applied in order to obtain diffraction-limited image sequences (i.e., around 0$\farcs$1 at H$\alpha$ and TiO). Figures~\ref{f1}(a) and (b) are the H$\alpha$ line center and offband images at the flare peak time, showing the double flare ribbons and the erupting material (pointed to by arrows). In fact, this flare is most probably preceded by the eruption of a filament lying along the PIL from $\sim$18:33:50 onward (see the time-lapse movie). A complete study of the erupting filament will be presented elsewhere. Figures~\ref{f1}(c) and (d) compare preflare and postflare TiO images, with the preceding spots P1--P3 and the following spot F1 labeled. It can be distinctly seen that after the flare, there is a newly formed penumbra (pointed to by the arrow) that directly connects P1 and F1 (see further discussion below). Contours of HMI line-of-sight (LOS) magnetic field is also over-plotted on the pre-flare TiO image in Figure~\ref{f1}(e) to clearly exhibit the polarity of each spot. BBSO did not have magnetograph observations on that day.

To shed light on the 3D magnetic field structure, we resorted to the nonlinear force-free field extrapolation (NLFFF), using HMI vector magnetograms as the boundary condition.  This extrapolation model uses the ``weighted optimization'' method as discussed by Wiegelmann (2004). The photospheric boundary is pre-processed to simulate the force-free condition (Wiegelmann et al. 2006). The calculation was performed to a volume of \sm171~$\times$~171~$\times$~171~Mm$^3$. Selected NLFFF lines are superimposed on Figures~\ref{f1}(e) and (f), demonstrating a possible sheared flux rope lying along the PIL (red) and the overlying arcade field (blue). These are nicely consistent with the above observation of the erupting filament and flare ribbons. In the standard flare picture, filament eruption can stretch open the overlying fields, which then reconnect and produce flare ribbons at their footpoints.

We present in Figure~\ref{f2} the time sequence of TiO images right across the flare (panels (a)--(g)) and a corresponding HMI vector magnetic field map (panel (h)). It is evident that there exists strongly sheared fields along the PIL between the positive, ridge-like P2 and negative F1 spots, and that before the flare, a sheared penumbra lies between P2 and F1 (Figure~\ref{f2}(a)). With the occurrence of the flare, a significant new section of penumbra then rapidly forms and is obviously seen to directly connect the main positive spot P1 with the negative F1 (Figures~\ref{f2}(e)--(g)) and hence link part of the flare ribbons (cf. Figures~\ref{f1}(b) and (d)). We note that some penumbral fibrils may apparently join positive spots P1 and P2. This is most probably because that the inclined penumbral field stemming from P1 could have a dip near the region of P2 but then curves up again to reach F1 (e.g., Figure 8 of Wang, Deng, \& Liu 2012). In the sense that the main spots P1 and F1 with opposite magnetic polarity now share a common penumbra, the flare effectively creates a new $\delta$ sunspot.

In order to further demonstrate the rapidness of the flare-related penumbral formation, we show in Figures~\ref{f3}(a) and (b) the temporal evolution of the mean TiO intensity within the red boxed region in Figure~\ref{f2} along with that of the corresponding horizontal magnetic field, in comparison with the GOES 1--8~\AA\ flux (the red line). It is remarkable that the TiO intensity begins to sharply drop with the onset of the flare at 18:45~UT and reaches a nearly 20\% lower level around 19:30~UT. At the meantime, the horizontal field starts to gradually increase from $\sim$18~UT but exhibits a highest increase rate (see the cyan line in (b)) simultaneously with the flare peak. Similar to our previous studies (e.g., Liu et al. 2005; Wang et al. 2012b), we believe that the rapid intensity darkening and horizontal field enhancement are resulted from the 3D magnetic field reconstruction due to the flare. The long-term gradual strengthening of the horizontal field, however, may be related to the overall evolution of the active region, i.e., the converging motion between P1/P2 and F1 as discussed below.

The spatial-temporal relationship of the formation of the new penumbra can be best seen in Figure 4, in which we show the time slice image along the slit as marked in Figure~\ref{f2}(a). Before the flare, obvious convective pattern is present in the upper region ($\sim$10--19\arcsec) with a time scale of 10--20 minutes and a spatial scale of about 1--2\arcsec, which are typical scales of granulation. After the flare, such a pattern is replaced by a typical penumbral structure with alternating dark and bright fibrils. It is also convincing that the transition from granulation to penumbrae is rapid and occurs right after the flare peak (the over-plotted solid line).

Since long-term evolution of magnetic field can provide hints for understanding the triggering of flares, we also examine in Figure~\ref{f3}(c) the time profiles of magnetic flux within the cyan box in Figure~\ref{f2}(h) that encompasses the main flaring PIL. We realize the difficulty in separating the short-term change in the smaller scale from the long-term change in the larger scale. Nevertheless, with the help of the accompanied magnetograph movie, there is some indication of magnetic cancellation in this region beginning from $\sim$18~UT, which reduces the positive flux with the highest cancellation rate (see the cyan line) co-temporal with the peak of the flare. The cancellation could be caused by the collision between P2 and the eastward motion of the emerging negative F1 field. However, it is also possible that such a reduction of positive flux is partially due to its southward migration. The evolution of negative field can not be used to identify the flux cancellation, as F1 is part of another larger scale flux emergence. The sharp increase of negative flux (see the green line) around the flare peak time is then due to the emerging flux moving into the calculation box. All these may be resulted from the interplay between the long-term and short-term magnetic restructuring, for which we presently can not draw a definite conclusion.

It is worth emphasizing two facts in our plots. First, a rapid change is only observed in the TiO intensity observed with NST. In contrast, all the HMI magnetic quantities have a more gradual change with an onset time even before the flare, although the sharp increase of the change rate is apparently associated with the flare. This could be attributed to the very different image scales of the two sets of observations and the long-term evolution of the entire active region in a large scale. Second, there is another peak in the field change rate around 21:15~UT, while we can not find any GOES X-ray signature around this time. After carefully examining the NST H$\alpha$ data, we noted a sub-flare peaked at 21:16~UT, which has a similar morphology as the C7.4 flare. We speculate that the magnetic field as well as intensity changes are associated with this sub-flare.

\section{SUMMARY AND DISCUSSION} \label{summary}

Taking advantage of the 0$\farcs$1 spatial resolution and 15~s cadence images of the NST, we observed in detail the rapid formation of a sunspot penumbra across the PIL closely associated with the 2012 July 2 C7.4 flare. The penumbral formation is unambiguously evidenced by the transformation from patterns of typical granulation to penumbral fibrils, and is accompanied with rapid TiO intensity darkening and horizontal field enhancement. The formation of this penumbra produces a larger scale $\delta$ sunspot. Such a fine-scale study is only possible with observations at sufficiently high resolution.

\begin{figure}
\epsscale{1}
\plotone{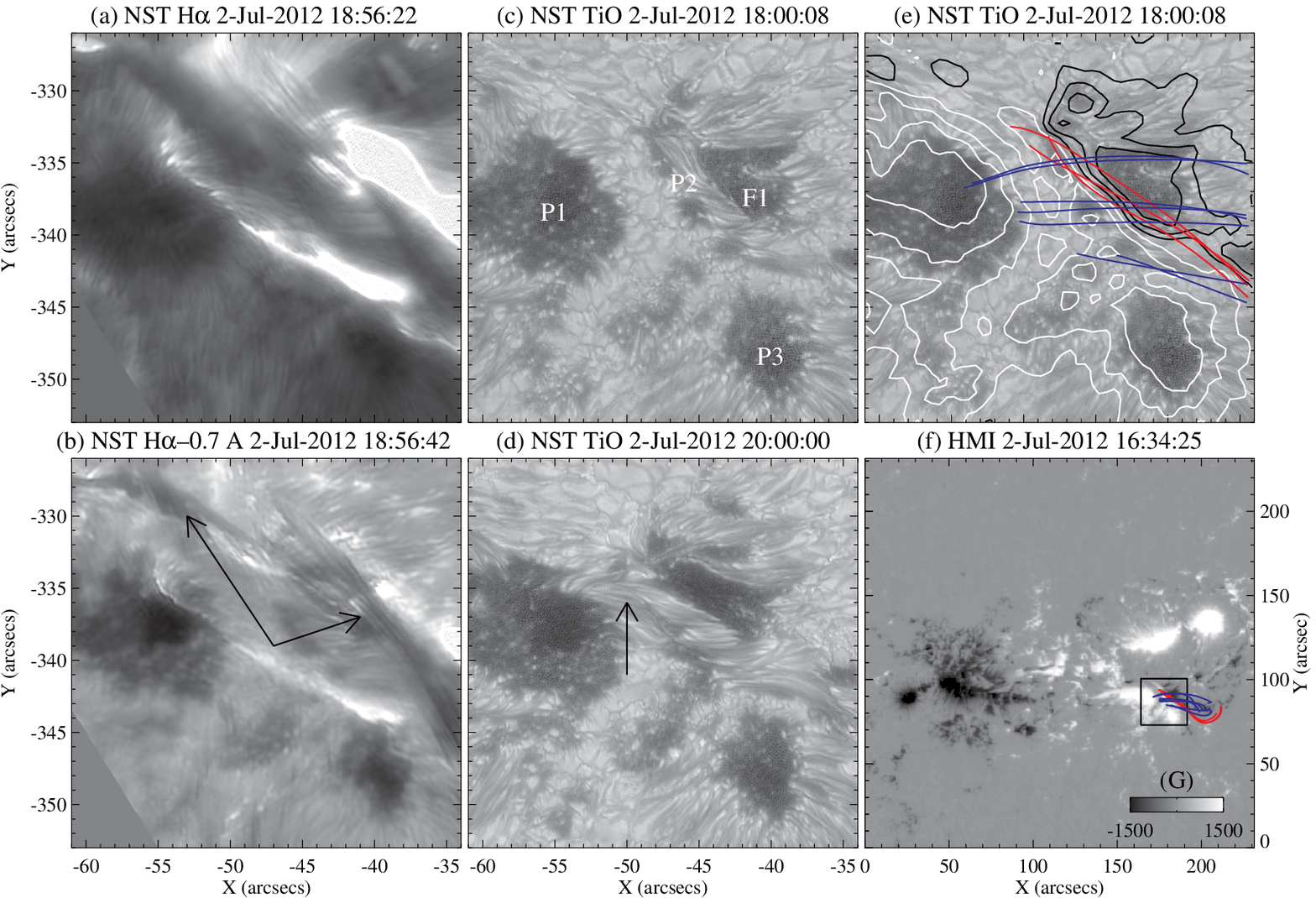}
\caption{NST \ha\ center (a) and blue-wing (b) images at the flare peak time showing the two flare ribbons and the possible signature of flux rope eruption (pointed to by the arrows in (b)). The NST TiO images about 1 hr before (c) and 1 hr after (d) the flare clearly show the formation of penumbra (pointed to by the arrow in (d)), which connect the northern two umbrae lying in the opposite magnetic field. The same preflare TiO image in (e) is superposed with positive (white) and negative (black) HMI LOS magnetic field contours, with levels of $\pm$200, $\pm$600, $\pm$1000, and 1400~G. All the images are aligned with respect to 18~UT. The remapped HMI vector magnetogram at 16:34~UT displaying the whole active region is shown in (f), overplotted with the black box denoting the approximately same FOV of (a)--(e). The superimposed red and blue NLFFF lines (also in (e)) represent the sheared flux rope along the PIL and the overlying arcade field, respectively. Two movies are associated with the Figure. movie-1a-ha-offband.mpeg is the H$\alpha$ blue wing movie, while movie-1b-tio.mpeg is the TiO movie.  \label{f1}}
\end{figure}

\begin{figure}
\epsscale{1}
\plotone{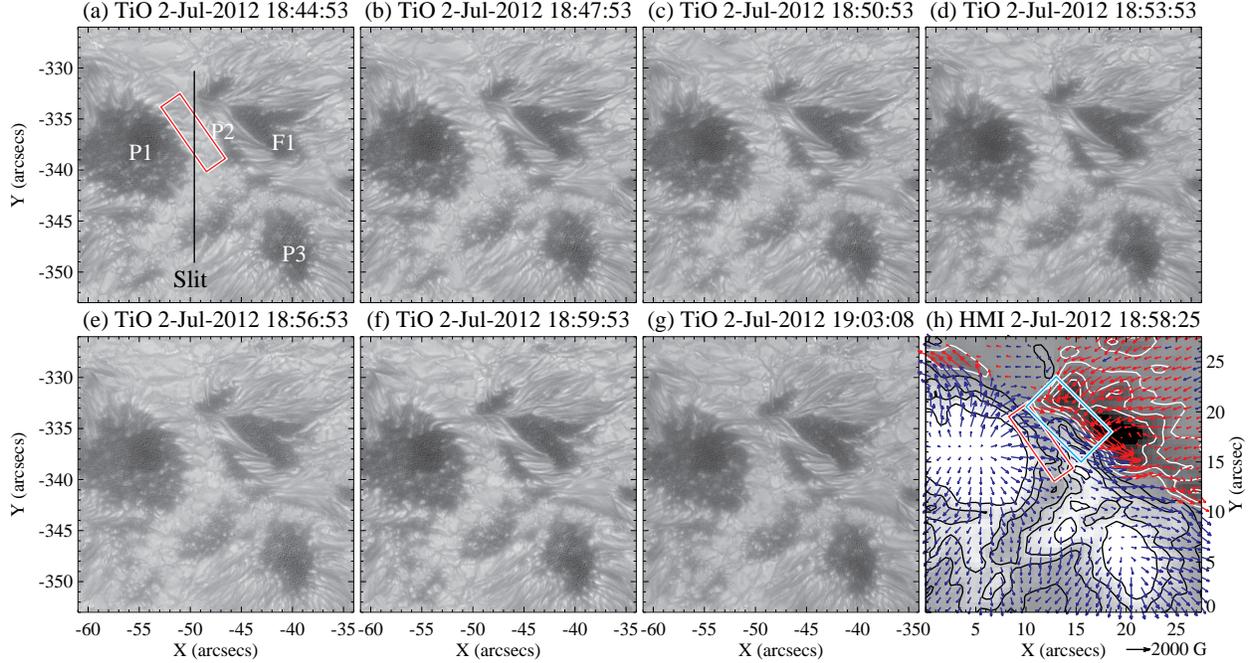}
\caption{(a--g) Time sequence of NST TiO images from right before to right after the flare, showing the gradual formation of penumbra. A remapped HMI vector magnetogram with an approximately same FOV is plotted in (h). The contours of vertical field have the same levels with those in Figure~\ref{f1}(e). The arrows represent horizontal magnetic field vectors. The red box overplotted on (a) and (h) is the region for which we calculate the temporal evolution of TiO intensity and horizontal magnetic field as shown in Figure~\ref{f3}(b). The cyan box overplotted on (h) is the region for which we measure the temporal evolution of vertical magnetic flux as shown in Figure~\ref{f3}(c). The black line in (a) is the slit using which we construct the time slices in Figure~\ref{f4}. \label{f2}}
\end{figure}

\begin{figure}
\epsscale{.65}
\plotone{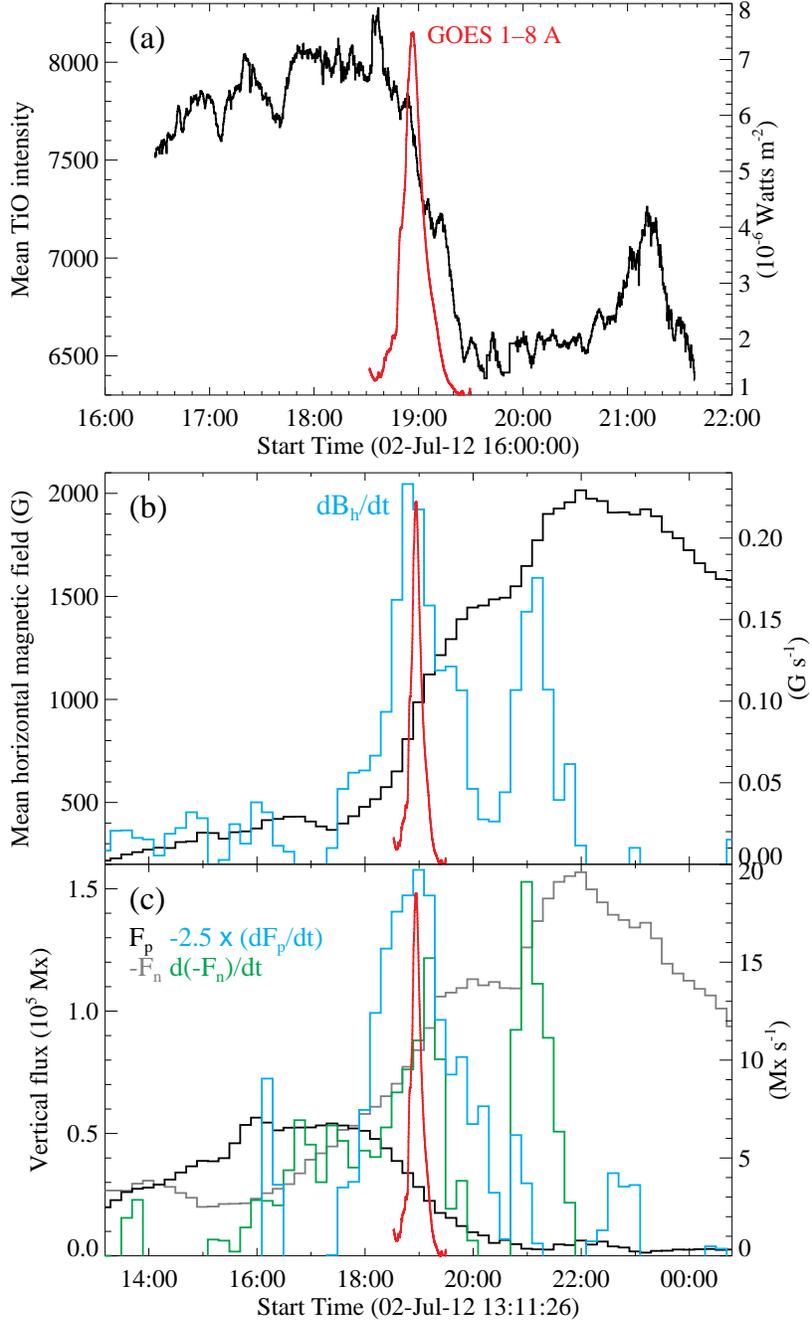}
\caption{Temporal evolution of the mean TiO intensity (a) and horizontal magnetic field (b) within the red boxed region in Figures~\ref{f2}(a) and (h), and the positive ($F_p$) and negative ($F_n$) magnetic fluxes (c) within the cyan boxed region in Figure~\ref{f2}(h). The overplotted red line is the GOES 1--8~\AA\ light curve for this flare. The colored lines in (b) and (c) are the temporal derivative of the corresponding quantities. Associated movie (movie-3-bl-bt.mpeg) shows the time sequence of longitudinal (left) and horizontal fields (right).  \label{f3}}
\end{figure}

Based on the observational results, we propose the following interpretation. There could exist a low-lying flux rope right above the initially sheared PIL, which can be supported by sheared arcade fields. Some evidence of magnetic cancellation at this PIL and the filament eruption preceding the flare are consistent with the formation and eruption of the flux rope (e.g., Green, Kliem, and Wallace 2010). As the flux rope moves upward, the closing of the opened overlying fields are then pushed back toward the surface as a result of the downward momentum (Fisher et al. 2012). This leads to the formation of the new penumbra and the $\delta$-spot configuration. Several supporting evidences for this scenario are as follows. (1) Sheared fields exist around the original PIL before the vertical fields cancel across it. (2) NLFFF extrapolation clearly shows both the flux rope and the overlying arcade fields.  (3) Rising of the flux rope is visible in the H$\alpha$ movie, and part of the bundle is apparent in the H$\alpha$ blue-wing image in Figure~\ref{f1}(b). (4) As shown in H$\alpha$ images (Figures~\ref{f1}(a) and (b)), the flare ribbons are not located beside the compact flaring PIL, but lie further apart in the main sunspots P1 and F1 that are finally connected by the newly formed penumbra. We also note that there are four homologous flares occurred during this period (17--22~UT). Each flare may have a partial contribution to the flux rope eruption but the C7.4 flare plays a major role. The penumbral formation is also only associated with this flare.

We emphasize that only high-resolution observing sequence would allow such a detailed examination of sunspot structure change. The size of the new penumbra is only about 5\arcsec. Without these new NST observations, we might only conclude on the feature darkening near the flaring PIL as what have been described in our previous observations (e.g., Liu et al. 2005). The formation of new penumbra also signifies the downward push of fields due to Lorentz force change, as implied by the more horizontal field lines after flares (Hudson et al. 2008; Fisher et al. 2012). As soon as some horizontal fields subside close to the surface, the penumbra can be formed. The opposite process was also observed before (Wang, Deng \& Liu 2012), in which the peripheral sunspot penumbra may disappear suddenly after flares, so that convection in the granule scale appears immediately as the fields turn from horizontal to vertical state. In both cases, penumbra may be quite a thin layer of structure, as discussed by Wang, Deng \& Liu (2012).

Although the study of larger scale flare activity is out of the scope of this Letter, it is worth mentioning that the present flare involves additional brightenings in other areas in the active region. In fact, by looking at Figure~\ref{f1}(f), we find a magnetic field configuration favorable for circular-ribbon flares (e.g., Pariat et al. 2010). That is, positive field surrounds the central negative field, and the negative field also exhibits obvious motion. Indeed, during this flare, EUV images taken by the Atmospheric Imaging Assembly (AIA) show a semi-continuous circular ribbon in the outer positive field together with the more compact ribbon in the central negative field region. The two ribbons in Figure~\ref{f1} as seen by NST represent part of the outer ribbon and most of the inner flare core. This circular flare picture is supported by the outward ejecting jets as clearly seen in the AIA movies. Flares having a circular ribbon have rarely been reported, although it is expected in the fan-spine magnetic topology involving reconnection at a 3D coronal null point (Masson et al. 2009; Reid et al. 2012; Wang \& Liu 2012). It is speculated that the eruption of flux rope may have been triggered by the outer null-point reconnection (e.g., Jiang et al. 2013), which cause homologous jets as observed by AIA.

\acknowledgments

We are grateful to the referee for the valuable comments to improve the paper. We thank the BBSO staff for obtaining the outstanding NST observations, and the SDO/HMI team for the magnetic field data. We are indebted to Dr. Yurchyshyn for pre-processing and transferring the NST data.  This work is supported by NSF grants AGS 1153226 and AGS 1153424, and NASA grants NNX13AG13G and NNX13AF76G.

\begin{figure}
\epsscale{1}
\plotone{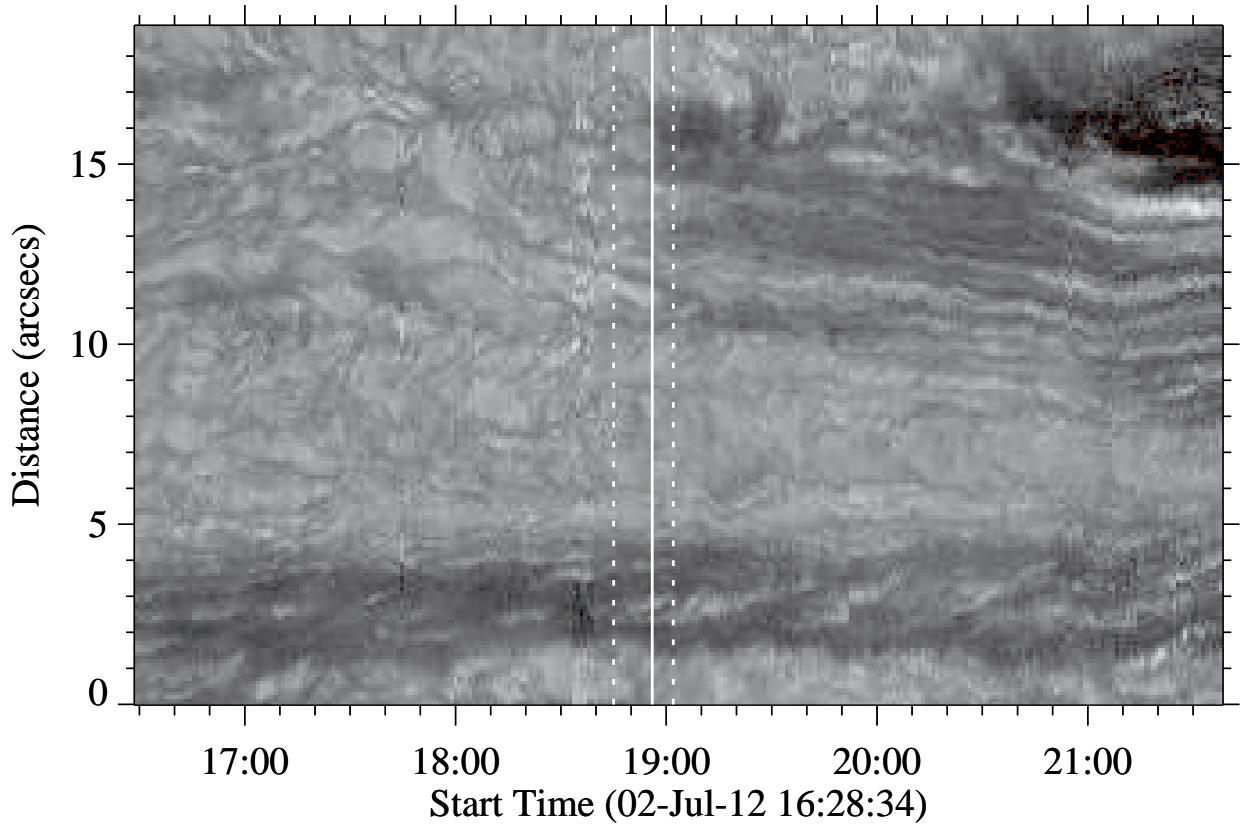}
\caption{TiO time slices for the slit shown in Figure~\ref{f2}(a). The distance is measured from the southern end of the slit. The dashed and solid lines denote the time of the start, peak, and end of the flare in GOES 1--8~\AA. \label{f4}}
\end{figure}

\end{document}